# PRODUCING HARD PROCESSES REGARDING THE COMPLETE EVENT: THE EPOS EVENT GENERATOR

S.Porteboeuf [*], T.Pierog [†], K.Werner

*SUBATECH, University of Nantes - IN2P3/CNRS - EMN, Nantes, France*

Jet cross sections can be in principle compared to simple pQCD calculations, based on the hypothesis of factorization. But often it is useful or even necessary to not only compute the production rate of the very high pt jets, but in addition the "rest of the event". The proposed talk is based on recent work, where we try to construct an event generator – fully compatible with pQCD – which allows to compute complete events, consisting of high pt jets plus all the other low pt particles produced at the same time. Whereas in "generators of inclusive spectra" like Pythia one may easily trigger on high pt phenomena, this is not so obvious for "generators of physical events", where in principle one has to generate a very large number of events in order to finally obtain rare events (like those with a very high pt jet). We recently developped an independnat block method which allow us ta have a direct access to dedicated variables [1]. We will present latest results concerning this approach.

## 1 Motivation : Jet production

High-$p_T$ Jets are rare processes which provide useful informations on the medium via jet-quenching [2]. Usually, when one wants to compute jets (hard processes), one uses the parton model, assuming factorization. But, one has to keep in mind here one only computes inclusive cross sections: P+P → Jet +X, with no possibility to investigate the "X-part". One cannot obtain partial cross sections, neither exclusive ones. In addition, the jet production is uncoupled from the rest of the event (soft production). As a consequence, event generators based on this approach could not compute complete event, they only compute jets, which may be supplemented by soft events from a different source.

The parton model hides multiples scatterings, which occur even in pp collisions. The importance of multiple scatterings was shown by [3], where the authors plotted the total cross section and the jet cross section as a function of the collision energy. For higher energies $\sigma_{jet} > \sigma_{Total}$. It means that there is more than one jet produced in a pp collisions, and therefore there is more than one interaction. As a consequence, if one wants to reproduce a complete event, an event that match the best what occurs in reality, one needs to consider multiples scattering.

In [4], the authors studied $p\bar{p}$ collisions at Tevatron energies, and they plotted the average $p_T$ as a function of the charged multiplicity. When the charged multiplicity increase, on get an events with increasing multiple interaction. What they observed is that the usual event generators where not able to reproduce the data. There is a real need of an event generator with a careful treatment of multiple scatterings. EPOS is actually such an generator of complete events, which generates hard scatterings in the context of multiple scattering. As a test, our approach must follow pQCD calculations for inclusive spectra at SPS and Tevatron energies. Some LHC prediction on our preliminary work can be found in [5].

---

[*] now at LLR, École polytechnique, Palaiseau, France
[†] Forschungszentrum Karlsruhe, Institut fuer Kernphysik, Karlsruhe, Germany

By computing jets in a complete event, we will have a real event generator. We want to do a correspondence :

$$1 \text{ experimental event } = 1 \text{ generator event} \qquad (1)$$

As a consequence, we will have a control over the underlying event. In other events generators, jets and soft part are computed in different manner : there is no connection between the two. Our approach gives us a better understanding of the event : jet regarding to the corresponding underlying event. Finally, a parallel work is done to implement an event generator with hydrodynamics [6]. For the future, on can expect an event generator with jet quenching, event by event.

## 2  Event generator : EPOS

EPOS stands for **E**nergy conserving quantum mechanical approach, based on **P**artons, parton ladders, strings, **O**ff-shell remnants, and **S**plitting of parton ladders). A compact description can be found in [7], many technical details about the physical basis of EPOS are described in [8], where we also discuss in detail the parameters of the model and how they are fixed. Concerning the basic features of this approach: EPOS is a consistent quantum mechanical multiple scattering approach based on partons and strings, where cross sections and the particle production are calculated consistently, taking into account energy conservation in both cases (unlike other models where energy conservation is not considered for cross section calculations). Motivated by the very nice data obtained by the RHIC experiments, nuclear effects related to Cronin transverse momentum broadening, parton saturation, and screening have been introduced into EPOS. Furthermore, high density effects leading to collective behavior in heavy ion collisions are also taken into account. It appears that EPOS does very well compared to RHIC data [9],[10], and also all other available data from high energy particle physic experiments (ISR,CDF and especially SPS experiments at CERN). As a result, EPOS is the only model used both for Extensive Air Shower simulations and accelerator physic which is able to reproduce consistently almost all data from 100 GeV lab to 1.8 TeV center of mass energy, including anti-baryons, multi-strange particles, ratios and pt distributions.

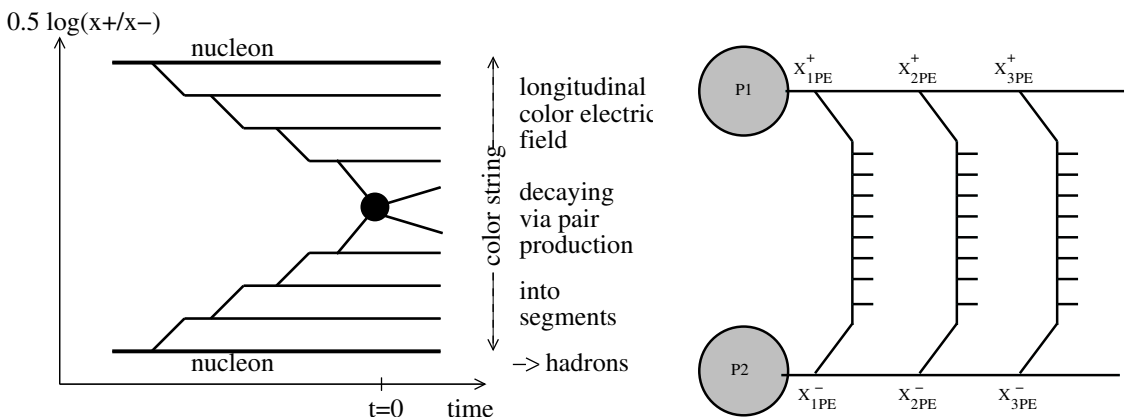

Figure 1: (left) Elementary Interaction. (right) Multiple Interaction : exchange of ladders in parallel, energy is shared between ladders.

The elementary interaction in EPOS is a parton ladder, see Fig.2 (left). The whole ladder could be seen as a longitudinal color electric field that decays via pair productions into seg-ments. Those segments are then identified as hadrons. Particle production follows the Lund String Model. In the initial stage of proton-proton collisions or heavy ion collisions, multiples interactions occur in parallel. This phenomenon represents in EPOS an exchange of parton ladder in parallel, see Fig.2 (right). In the two limits, a parton ladder could either be soft or hard. This representation allows us to consider semi-hard cases. What is important to remark is that both, hard and semi-hard interactions are computed in the same formalism. Finally, this picture produces hard scattering regarding the complete event.

One can see on Fig.2 (right) a scheme of an event with multiple interaction. Several ladders are exchanged in parallel. All ladders exist at the same time. The total energy is shared between ladders. This means that it's impossible to have an infinite number of ladder. The total energy is conserved. $x_{PE}$ refers to the fraction of light cone momentum of the parton which enter the ladder.

## 3 Hard process in a complete event

We now zoom on one ladder. This ladder can be soft – with a parametrization in regge fashion – or it can be be semi-hard, as shown in Fig.2. In the semi-hard case, there is a soft pre-evolution, followed by a DGLAP evolution. In Fig.2, the two evolutions are contained in the central blob. Finally, in the center of the ladder, there is a hard process $2 \rightarrow 2$ (green part of Fig. 2).

Once the Monte Carlo procedure has determined how many ladders there are, and the $x_{PE}$ values at the ladder ends, we employ an iterative procedure to compute emission up to the hard scattering $2 \rightarrow 2$. We define $x_{IB}$ as the light cone momentum entering the hard process, while $x_{PE}$ refers to the light cone momentum entering the ladder. For complete definition of variables, sea section ??. One first generates all resolvable partons emitted at one side of the ladder before the hardest process. At each step one decides whether there is any resolvable emission at the forward end of the ladder before the hardest process. In case of no forward emission, the generation of all resolvable parton emissions at the forward side of the ladder has been completed. One then proceeds to generate all resolvable parton emissions for the backward side of the ladder by using a corresponding recursive algorithm. For more details about probability of forward and backward emission and the probability distribution of the light cone momentum and momentum transfer squared for the current parton branching, see [8].

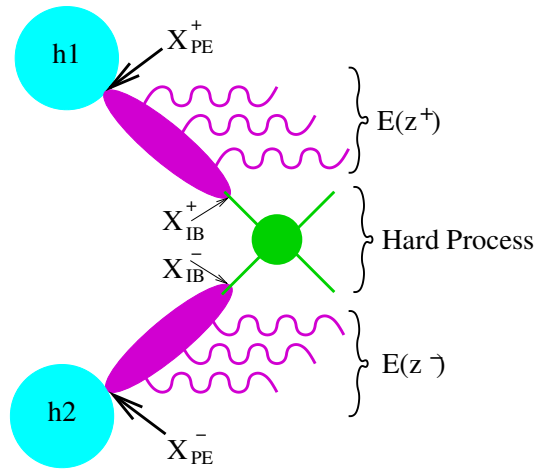

Figure 2: A semi-hard ladder

In the usual event generators based on parton model are generators of inclusive spectra, as discussed earlier. In such generators, one can easily work with cuts, in order to look at rare event such as very high $p_T$ jets. In our case of a generator of "complete events" this is more complicated. An inconvenience is that, like in experiments, if one wants rare event, one needs a a large number of simulations. The question is now : How to compute hard partons, in a generator with multiple scattering, with cuts (to look at rare events)?

In[1], we presented the selection cut method based on a two-step procedure, where the internal part of the ladder is rewritten with independant block. A detailed description of the independant block method can be found in[11]. We will now present the latest results concerning this approach.

## 4 Computation of hard parton cross section

In this section, we will describe observables from the EPOS framework that can be compared to measurment, the inclusive jet cross section. To do so, we need to express the internal quantity $n_{\text{semi}}$, which represents the number of semi-hard ladder (ladders describe by Fig. 2), as a function of $p_T$. This expression is expressed here in for the case of factorisation, which araise naturally in the EPOS framework[8]. This limit describe inclusive spectrum such as the inclusive jet cross section. This observable is a good first test for the independant block method in the limit of factorisation. The independnat block method can then be used to construct a new Monte Carlo with the implementation of cuts.

$$\frac{dn_{\text{semi}}}{dtdu} = \int dx_{\text{IB}}^+ dx_{\text{IB}}^- \sum_{ij} f_+^{M,i}(x_{\text{IB}}^+) f_-^{M,j}(x_{\text{IB}}^-) K_{ij}(x_{\text{IB}}^+ x_{\text{IB}}^- s, t, u) \qquad (2)$$

In eq.2, $n_{\text{semi}}$ is expressed as a function of the internal variables of the ladder, described in Fig. 2. $K$ represents the internal block of the $2 \to 2$ process and $f$ the evolution of the parton from the proton to $2 \to 2$ process. A more complete description can be found in [1,11].

With:
$$dtdu = |\frac{\partial(t,u)}{\partial(y,p_\perp)}|dydp_\perp = \frac{2tu}{p_\perp}dydp_\perp = sdydp_\perp^2 \qquad (3)$$

we obtain:

$$\frac{dn_{\text{semi}}}{dydp_\perp^2} = \int dx_{\text{IB}}^+ dx_{\text{IB}}^- \sum_{ij} f_+^{M,i}(x_{\text{IB}}^+) f_-^{M,j}(x_{\text{IB}}^-)$$
$$\times sK_{ij}(x_{\text{IB}}^+ x_{\text{IB}}^- s, t) \delta(x_{\text{IB}}^+ x_{\text{IB}}^- s - x_{\text{IB}}^+ p_\perp \sqrt{s} e^{-y} - x_{\text{IB}}^- p_\perp \sqrt{s} e^y)$$

$$\frac{dn_{\text{semi}}}{dydp_\perp^2} = \frac{1}{\sigma_{\text{inel}}} \int dx_{\text{IB}}^+ \sum_{ij} f_+^{M,i}(x_{\text{IB}}^+) f_-^{M,j}(\tilde{x}_{\text{IB}}^-) s \frac{d\sigma_{ij}}{dt}(s,t) \frac{s_{hh}^{-1}}{x_{\text{IB}}^+ - x_\perp e^y/2} \qquad (4)$$

with:

$$s = x_{\text{IB}}^+ x_{\text{IB}}^- s_{hh}, \; \tilde{x}_{\text{IB}}^- = \frac{x_{\text{IB}}^+ x_\perp e^{-y}/2}{x_{\text{IB}}^+ - x_\perp e^y/2}, \; x_\perp = \frac{2p_\perp}{\sqrt{s_{hh}}}$$

Finally, if one wants to compute the number of outborn partons (partons from the $2 \to 2$ process that initiate jets), one need to multiply by two the formula, each ladder gives two hard partons: $n_{\text{ptn}} = 2n_{\text{semi}}$.

## 5 Results

The test of the approach is presented in Fig.5, where we compare EPOS analytic (facorization case) results from section 4 to experimental data for the inclusive jet spectrum in proton-proton collision at $\sqrt{s} = 200$ GeV. On the left hand-side the spectrum is presented and compare with results from cross-section compiled with PDF (GRV and CTEQ6) and an NLO computation with variable cone radius of the jet. Data are exctracted from [12]. There is two data sets, each one being a combination of 2003 and 2004 runs. Empty square refers to a minimum bias selection and full point refers to a high tower trigger selection: event where there is an energetic jet with an energetic leading particle. Jet are identified with a cone jet-finder algorithm with a cone radius of $R = 0.4$, chosen in regards of the acceptance of the detector, according to the author, 95% of the total energy of the jet is expected to be in a cone of a radius of 0.4.

We can see that EPOS is close to STAR data with less than a factor two over nine orders of magnitude. This is also shown an the right hand-side of Fig. 5, upper plot, where ratio of STAR over EPOS computation is plotted.At high $p_T$ EPOS is in between the two PDF sets. Data are also compared with NLO QCD computation from [13]. This specific contribution is able ta take into account the variation of the cone radius of the jet. Here we want to point out the fact that in this EPOS compuation, we compute the production of a hard parton, which is, in practise, not exaclty the same observable as a reconstructed jet. To go furter into details, one should compare event produced inside the event generator with this independant block method. This work is in progress. Finaly this NLO QCD computation should be in better agreement with the data than the 3 other computation, the ratio of STAR/NLO is shown on the right-hand side, middle plot. The lower plot shows ratio of EPOS over NLO, which is, at first order, in good agreement. Finally, it appears that the comparison to data is a first good test fo the independant block method.

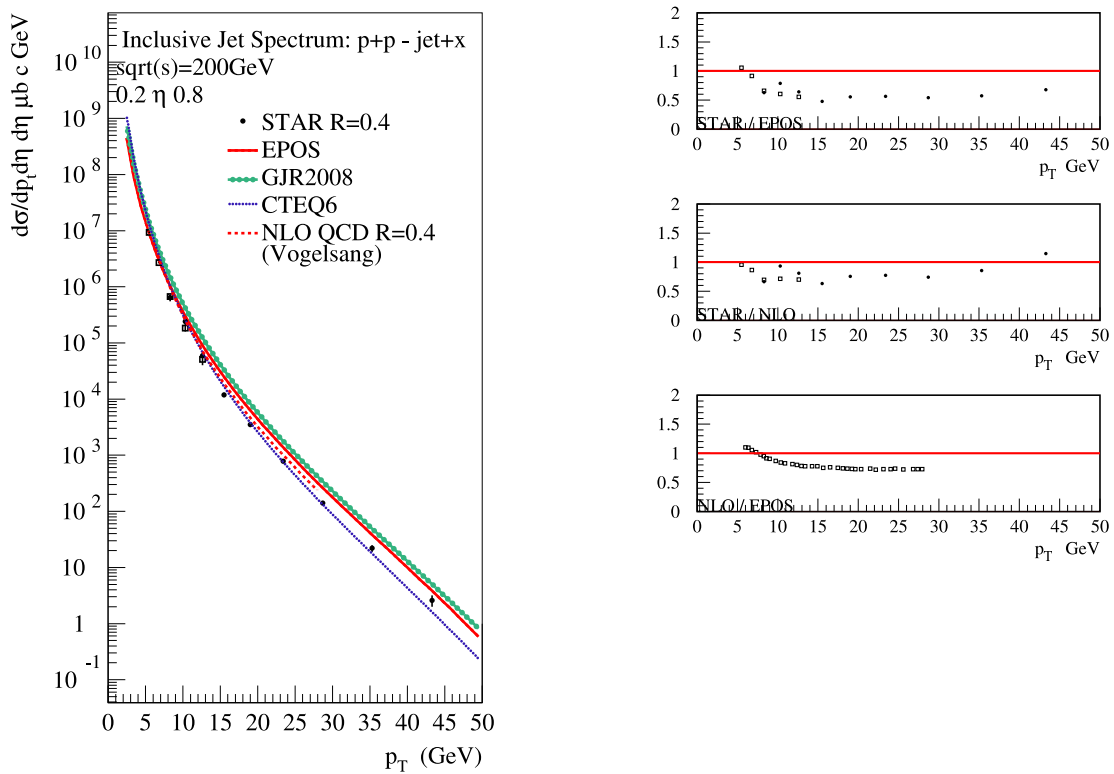

Figure 3: Inclusive jet cross section for pp → Jet+X for $\sqrt{s} = 200$GeV. (left) Spectrum for Star data, EPOS, computation cross section with PDF set, GJR2008 and CTEQ6 and NLO QCD from. (right) ratios : upper plot STAR/EPOS, middle plot: STAR/NLO, lower plot: NLO/EPOS.

## 6  Conclusion

EPOS has been constructed as an event generator of complete events, very close to an experimental event, with soft and hard parts in the same formalism. To produce complete event, we consider multiple scattering with energy conservation, and we compute hard partons in this context. We are able to produce a jet connected to its event. One benefit is to control the underlying event and to connect an underlying event to the production of a jet. The price to pay is that – as in experiments – high $p_T$ jets are rare and one needs many simulations to observe one. Therfore we present a procedure to produce hard partons by making use of cuts, to reasonable statistics in a reasonable time. The method as describe earlier is to change the procedure for computing the internal part of the ladder. A validation of the method is shown by the comparison of EPOS computation with STAR data for the inclusive jet spectrum for proton-proton scatering at $\sqrt{s} = 200$ GeV. This comparison validate the independant block method, and encourage the developement of the new approach into the compete event procedure. This work is in progress.